\documentclass[preprint,showpacs,preprintnumbers,amsmath,amssymb]{revtex4}

\usepackage{graphicx}
\usepackage{dcolumn}
\usepackage{bm}

\begin{document}


\title{Magnetic state dynamics in itinerant paramagnet U$M_{3}$B$_2$\\
($M$ = Co, Ir) probed by $^{11}$B NMR}

\author{Tatsuya Fujimoto$^{1}$}
 \email{fujimoto.tatsuya@jaea.go.jp}
\author{Hironori Sakai$^{1}$}
\author{Yo Tokunaga$^{1}$}
\author{Shinsaku Kambe$^{1}$}
\author{Russell E. Walstedt$^{1}$}
\author{\\Shugo Ikeda$^{1}$}
\author{Jun-Ichi Yamaura$^{2}$}
\author{Tatsuma D. Matsuda$^{1}$}
\author{Yoshinori Haga$^{1}$}
\author{Yoshichika \={O}nuki$^{3,1}$}
\affiliation{
$^{1}$Advanced Science Research Center, Japan Atomic Energy Agency, Tokai-mura, Ibaraki 319-1195, Japan\\
$^{2}$Institute for Solid State Physics, University of Tokyo, Kashiwa, Chiba 277-8581, Japan\\
$^{3}$Graduate School of Science, Osaka University, Toyonaka, Osaka, 560-0043, Japan
}%

\date{\today}

\begin{abstract}
We have carried out the $^{11}$B NMR measurement on the itinerant paramagnetic systems U$M_{3}$B$_{2}$ ($M =$ Co, Ir)
to investigate the low-dimensional characteristics of the $5f$-electrons due to the structural anisotropy.
The recent X-ray analysis suggests that
UIr$_3$B$_2$ has a different structure modulated from the ever-known superlattice.
The azimuth angle variation of NMR spectrum within the $ab$-plane clarified that B atoms occupy the single site, and
a certain ligands arrangement surrounding B atom turns to the same orientation as the another one through the three- or six-fold rotation around the $c$-axis.
These results have been consistent with the X-ray proposition.
To evaluate the temperature ($T$) development of general susceptibility ($\chi_{q,\omega}$),
Knight shift and nuclear spin-lattice relaxation rates measurements were performed and
the similar variations of $\chi_{q,\omega}$ were identified in both UCo$_{3}$B$_{2}$ and UIr$_{3}$B$_{2}$.
Above a crossover point defined as $T^{*}\simeq50$ K,
the evolution of $\chi_{q,\omega}$ is dominant at $q=0$,
suggesting that ferromagnetic correlations develop in high-$T$ regimes; meanwhile, below $T^{*}$,
the $q=0$ part in $\chi_{q,\omega}$ shows the saturation tendency, and
a different class of dispersion at finite-$q$ suddenly emerges.
This particular magnetic correlations are interpreted as the antiferromagnetic correlations, and
notable feature of the magnetic state dynamics in low-$T$ regimes is
that the antiferromagnetic correlations arise together with the ferromagnetic component at the same time.
The unique magnetic correlations obtained from NMR experiment will be discussed
by the possible low-dimensionality of U$M_{3}$B$_{2}$ lattice.
\end{abstract}

\maketitle

\section{\label{introduction}Introduction}

The low-dimensional electron and spin systems have attracted much interest for a long decade,
because these systems show a rich variety of physical properties.
Highly anisotropic electron state and magnetic exchange lead a peculiar instability and quantum fluctuations.
For example, in one-dimensional electron systems,
insulating charge density wave arises due to the so-called Peierls instability of electron-phonon interaction,\cite{peierls} or otherwise
the anomalous conducting phenomena due to the critical fluctuations is well-known as Tomonaga-Luttinger liquid.\cite{tomonaga,luttingere}
Another representative example in spin systems will a resonating valence bond state on two-dimensional triangular lattice,\cite{RVB} and 
several experimental attempt to prove the theoretical predictions has been performed in the past year. (e.g., Ref. \onlinecite{RVB-exp1})

Many works investigating low-dimensional properties have focused on the transition metal oxides and organic compounds,
because a lot number of model substances has been found in these systems. 
On the other hand,
$f$-electron intermetallics, especially heavy-fermion systems, have been given rather less attentions.
Recently,
single crystal studies of ternary lanthanide bolides $Ln$Rh$_{3}$B$_{2}$ ($Ln=$Ce, Pr) have been performed
by T. Okubo and M. Yamada \textit{et al}., and
these compounds have been regarded as low-dimensional electrons and magnetic systems.\cite{CeRh3B2,PrRh3B2}

\begin{figure}
\includegraphics{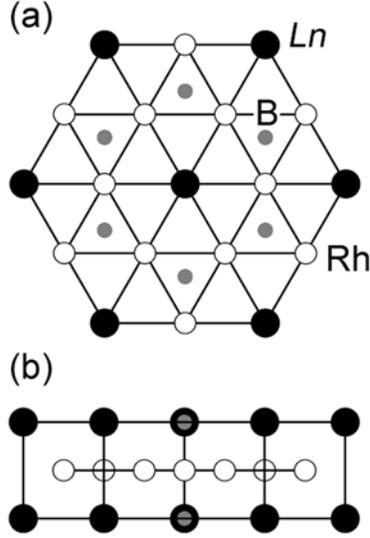}
\caption{\label{structure}
The projection of $Ln$Rh$_{3}$B$_{2}$ ($Ln=$Ce, Pr) structure onto (a) $ab$-plane and (b) $ac$-plane.
In this unit cell, the lattice constant $c$ is notably shorter than $a$-length.
The closed, open and shadow circles mean $Ln$, Rh and B sites, respectively.
All of the elements are crystallographically single-site, and
$Ln$, Rh and B forms the triangular, kagom\'{e} and honeycomb sublattices, respectively.
}
\end{figure}

The structural lattice of $Ln$Rh$_{3}$B$_{2}$ is displayed in Fig. \ref{structure}.
These compounds crystallize into hexagonal CeCo$_{3}$B$_{2}$-type structure with space group P6/mmm.
In this structure, two considerable feature,
which will frequently affect on the physical properties of many-body electron systems,
are contained.
The first feature is the one-dimensionality along the $c$-axis.
The lattice constant along the $c$-axis is markedly shorter than $a$-axis (for example, $a=5.469 \textrm{\AA}$ and $c=3.096 \textrm{\AA}$ for CeRh$_{3}$B$_{2}.$\cite{Struc1,Struc2,Struc3})
, so that
the strong correlations and the good conductivity along the $c$-axis will be reminded in a series of these compounds.
From the comparison between de Hass-van Alphen experiment and band calculation,
a wavy but almost flat Fermi surface concerned to the one-dimensionality has been actually concluded in both CeRh$_{3}$B$_{2}$ and PrRh$_{3}$B$_{2}$,\cite{CeRh3B2,PrRh3B2}
and correspondingly,
the relatively larger magnitude of resistivity along the $c$-axis than the $ab$-plane has been clarified.\cite{PrRh3B2}
The another striking feature of this compounds is a geometrical frustration.
As shown in Fig.~\ref{structure},
$Ln$ sublattices form the triangular lattice, and
ferrimagnetism in PrRh$_{3}$B$_{2}$, originating from an Ising-type classical antiferromagnetic frustration has been suggested.\cite{PrRh3B2}

These results in lanthanide systems have been encouraged us to develop into the actinide families U$M_3$B$_2$ ($M=$Co, Ir),
and we have undertaken $^{11}$B NMR measurement to look into low-dimensional characteristics of $5f$-electron systems.
The UCo$_{3}$B$_{2}$ crystallizes into the isomorphic $Ln$Rh$_3$B$_2$,
and the UIr$_3$B$_2$ forms the modulated superlattice due to the slight atomic displacement from the CeCo$_3$B$_2$-type structure.\cite{yamaura}
Since the structural difference between these is insignificant,
systematic comparison of the observed results can be possible.
These compounds displays the itinerant paramagnetic behavior,\cite{previous1,previous2,previous3,ikeda} and
no magnetic ordering have been observed down to 0.5 K.\cite{ikeda}
In the low temperature ($T$) state,
the resistivity obeys the Landau-Fermi liquid type variations, and
the Kadowaki-Woods law exhibited an amount of Sommerfeld coefficient ($\gamma$) as a few hundred mJ/mol$\cdot$K$^{2}$ for UIr$_{3}$B$_{2}$ and a tenth order for UCo$_{3}$B$_{2}$.\cite{ikeda}
The larger $\gamma$ in UIr$_3$B$_2$ than in UCo$_3$B$_2$ suggests that
UIr$_{3}$B$_{2}$ has more enhanced quasi-particle band on Fermi level.
In connection with this matter,
the resistivity of UIr$_{3}$B$_{2}$ draws the shoulder-like curve which is the signature of dense-Kondo systems.
On the other hand,
simple metallic behavior has been observed in UCo$_{3}$B$_{2}$.\cite{ikeda}

In this paper,
we report the $^{11}$B NMR results investigating the crystallographic and electric properties in U$M_3$B$_2$.
The NMR spectrum and its azimuth angle ($\theta$) variation in UIr$_3$B$_2$ indicate that B site is single, and
a ligands orientation surrounding B atom points the same way as the another one by three- or six-fold rotation around the $c$-axis.
This NMR results obtained have been consistent with the modulated superlattice proposed by X-ray diffraction measurement.\cite{yamaura}
The Knight shift and nuclear spin-lattice relaxation rates
varies its behavior around $T^{*}\simeq50$ K in both UIr$_3$B$_2$ and UCo$_3$B$_2$,
which implies that the crossover of magnetic state dynamics commonly occurs in these compounds at the same $T$.
Moreover,
similar magnetic correlations have been identified in each systems from the evaluation of general susceptibility ($\chi_{q,\omega}$).
Above $T^{*}$,
the development of $\chi_{q,\omega}$ is mainly caused on $q=0$,
suggesting that ferromagnetic correlations grown in high-$T$ regime.
Below $T^{*}$,
the ferromagnetic components in $\chi_{q,\omega}$ still maintain its magnitude, in addition,
the finite-$q$ dispersions in $\chi_{q,\omega}$ standing for the antiferromagnetic correlations suddenly emerge.
The origin of simultaneous observation of ferromagnetic and antiferromagnetic correlations in both U$M_3$B$_2$
is discussed on account of the low-dimensionality associated from the lanthanide $Ln$Rh$_3$B$_2$.\cite{CeRh3B2,PrRh3B2}
The part of the works in UIr$_{3}$B$_2$ has already been published on previous proceedings,\cite{fujimoto1,fujimoto2}
where experimental results and its interpretations have been briefly stated.
In this paper,
full set of data including UCo$_3$B$_2$ will be presented, and
qualitative analysis and comparison between two intermetallic bolides are carried out.

\section{\label{experiment}Experiment}

The polycrystalline UCo$_3$B$_2$ and the single crystal UIr$_3$B$_2$\cite{twin} were grown by the Czochralski pulling method in tetra-arc furnace.
The details of sample preparation will be described in elsewhere.\cite{ikeda}
For the NMR measurement,
UCo$_3$B$_2$ was powdered in order to penetrate the rf pulses sufficiently, and
UIr$_3$B$_2$ was sliced into $1\times1\times1$ mm$^{3}$ dimensions for the easy treatment.
The natural line width of these compounds were narrow ($\simeq10$ kHz),
indicating that the samples used in this paper have a homogeneous electron state without any lattice stains.

NMR measurements were carried out by conventional spin-echo method with $90^{\circ}$-$180^{\circ}$ series of pulse,
using phase-coherent pulsed-spectrometer.
$T$-range is 5-270 K, and
applied magnetic field ($H$) was fixed to be $H\simeq8$ T for UCo$_3$B$_2$ and $H\simeq7$ T for UIr$_3$B$_2$.
The NMR spectrum was obtained from the Fast Fourier-transformation technique.
The Knight shift of UIr$_3$B$_2$ ($K_\alpha$; $\alpha=ab, c$) were measured at the peak-frequency of central resonance ($+1/2\leftrightarrow-1/2$ transition)
within the $ab$-plane (parallel to $[210]$) and $c$-axis (parallel to $[001]$).
Meanwhile,
the Knight shift of UCo$_3$B$_2$ ($K_{\textrm{av}}$) was estimated from the gravity of central resonance, and
$K_{\textrm{av}}$ means the spatial average of generally anisotropic Knight shift.
This comes from the powdering procedure.
The nuclear spin-lattice relaxation rate of UIr$_3$B$_2$ ($[T_1T]_\alpha^{-1}$) and UCo$_3$B$_2$ ($[T_1T]_\textrm{av}^{-1}$)
was obtained from the saturation-recovery method ($90^{\circ}$-$90^{\circ}$-$180^{\circ}$ pulse sequences), and then,
satellite resonance ($-1/2\leftrightarrow+1/2$ transitions) for UIr$_3$B$_2$ and all of the spectrum components for UCo$_3$B$_2$ was excited.
Obtained nuclear-magnetization recoveries was uniquely determined by a single components function
with the magnetic relaxation process\cite{Narath} over the measured $T$-range.
This means that the obtained amount of nuclear spin-lattice relaxation rate and its $T$-variations are accurate and reasonable.

\begin{figure}
\includegraphics{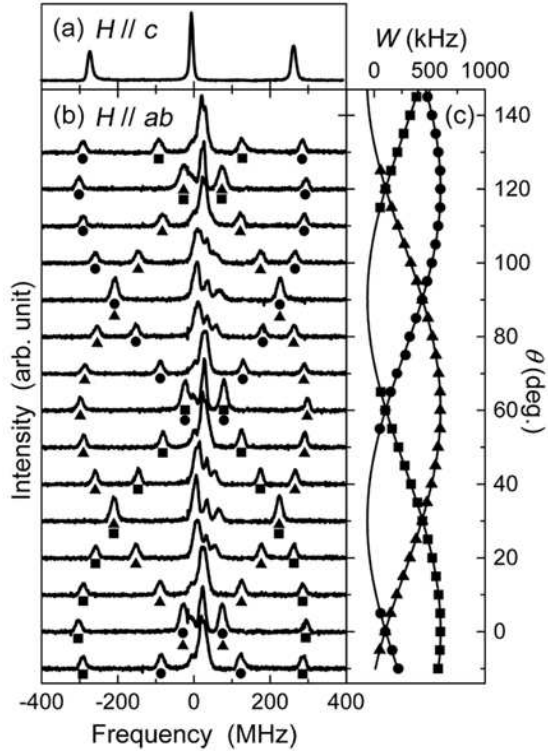}
\caption{\label{UIr3B2-spec}
The $^{11}$B NMR spectrum of UIr$_3$B$_2$, (a) along the $c$-axis and (b) within the $ab$-plane.
In (b), the azimuth angle ($\theta$) between $H$ and [210] are scaled in right hand longitudinal axis.
The three different pairs of satellite are denoted as circles, squares and triangles, respectively.
The $\theta$-variations of satellite spacing ($W$) are plotted in (c).
The solid lines are the fit results described in text.
}
\end{figure}

\section{\label{Experimental Results}Results and Discussions}
\subsection{\label{spectrum}NMR spectrum}

\begin{figure}[b]
\includegraphics{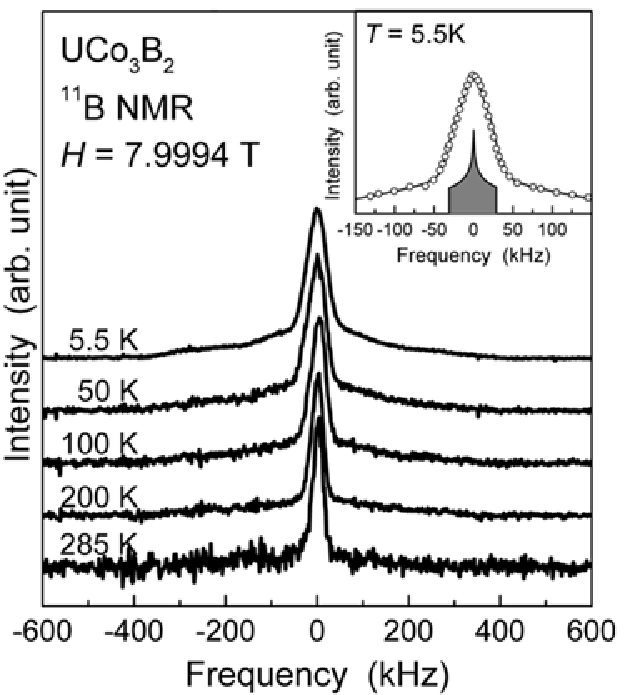}
\caption{\label{UCo3B2-spec}
The $^{11}$B NMR spectra of UCo$_3$B$_2$ powder at several temperatures.
The inset is the magnified figure of central-resonance at 5.5 K, and
the solid line is an outcome from the general powder-pattern (shadow part) convoluted with the Lorentz functions.
}
\end{figure}

Since the crystal structure of UIr$_3$B$_2$ does not still determined precisely,
we measured the NMR spectrum in several orientation to mention the crystal symmetry.
The Fig. \ref{UIr3B2-spec}(a) and \ref{UIr3B2-spec}(b) shows the $^{11}$B NMR spectra of UIr$_3$B$_2$ at 5.5 K.
In the case of $H\parallel c$,
a simple NMR spectrum with the small electric field gradient (EFG) on $I=3/2$ nuclear spins have been observed.
This result concludes that the magnitude of EFG is the same on all $^{11}$B nuclei, and thus
crystallographically single site is inferred for B atoms.
The $H$-application within the $ab$-plane splits NMR spectrum
into the three sub-spectra.
The ratio of three sub-spectra is $1:1:1$.
From the $\theta$-variation of NMR spectrum within the $ab$-plane,
the quite similar behaviors of satellite spacing ($W$) for each sub-spectra was found;
whose curves trace the same amplitude and the periodicity, but its phase alternates $60^{\circ}$ in orders.
The $z$-axis of EFG tensor points to [210] ($\theta=0^\circ$) or its equivalent directions,
suggested by the six-fold symmetry of NMR spectrum and the larger $W$-maxima than that of $c$-axis.
This results represent that
a certain ligands orientation enclosing B atoms can have the same arrangement as the another one
by the $60^{\circ}$ or the $120^{\circ}$ sample rotation around $c$-axis.
The recent proposed structure by the X-ray diffraction measurement\cite{yamaura} can explain the rotational variation of NMR spectrum, and
The details of X-ray analysis will be appeared in separate paper.

To evaluate an amount and anisotropy of EFG,
we estimate the nuclear quadrupole frequency ($\nu_\textrm{Q}$) and asymmetric parameter ($\eta$)
defined as $\nu_\textrm{Q}=3e^2QV_{\textrm{zz}}/20\pi \hbar$ and $\eta=(V_{\textrm{yy}}-V_{\textrm{xx}})/V_{\textrm{zz}}$,
where $Q$ is the nuclear quadrupole moments of $^{11}$B nucleus,
and $V_{\beta\beta}$ ($\beta=x,y,z$) is the magnitude of EFG along the $\beta$-axis.
These parameters can be obtained from the fitting of $W$ with the theoretical function described as,\cite{W}
\begin{equation}
W=\nu_\textrm{Q}(3\textrm{cos}^{2}\theta -1-\eta \textrm{sin}^{2}\textrm{cos}2\phi).
\label{nu_Q-eta}
\end{equation}
In the case of $\phi=0^{\circ}$ (therefore $V_\textrm{yy}\parallel c$),
the fairly fit could not be done due to the theoretical demand on $V_{\textrm{xx}}\leq V_{\textrm{yy}}\leq V_{\textrm{zz}}$.
As a result,
$V_\textrm{xx}$ is determined along the $c$-axis and fitting was performed with $\phi=90^{\circ}$.
The results are displayed in Fig. \ref{UIr3B2-spec}(c) as the solid lines.
Although the fit has been carried out independently for three different $W$,
the almost same $\nu_\textrm{Q}$ and $\eta$ was obtained.
This is the another proof of crystallographically single B site.
The mean value of EFG parameters are $\nu_\textrm{Q}=300$ kHz and $\eta=0.79$.
The rather large $\eta$ would imply that
Ir ligands surrounding B atom distorts from the CeCo$_3$B$_2$-type structure, in which
axial EFG symmetry ($\eta=0$) is expected,
because B atoms locate on the gravity of regular prismatic-Ir and U-triangle.

Fig. \ref{UCo3B2-spec} shows the $T$-variation of $^{11}$B NMR spectra in UCo$_3$B$_2$.
The satellite resonances were broadened due to the sample powdering process.
The satellite components in NMR spectrum spread over the $\pm400$ kHz,
and this amount is comparable to the $\nu_\textrm{Q}$ obtained in the UIr$_3$B$_2$ (300 kHz).
Since the approximately identical valence is assumed in both compounds,
the structural similarity between these will cause the same order of $\nu_\textrm{Q}$.
The central resonance is also broadened due to the anisotropic Knight shift,
and we performed the analysis of the obtained spectrum at 5.5 K with general powder pattern.\cite{PP}
The result is shown in the inset of Fig. \ref{UCo3B2-spec}.
Since the peak and two mid-points in central resonance correspond to the each Knight shifts along its principal axis.
these mean ($K_{\textrm{av}}$) can be interpreted as isotopic part, and
its $T$-variations will be discussed in next subsection.

\subsection{\label{shift}Knight shift and $K-\chi$ plot}

\begin{figure}
\includegraphics{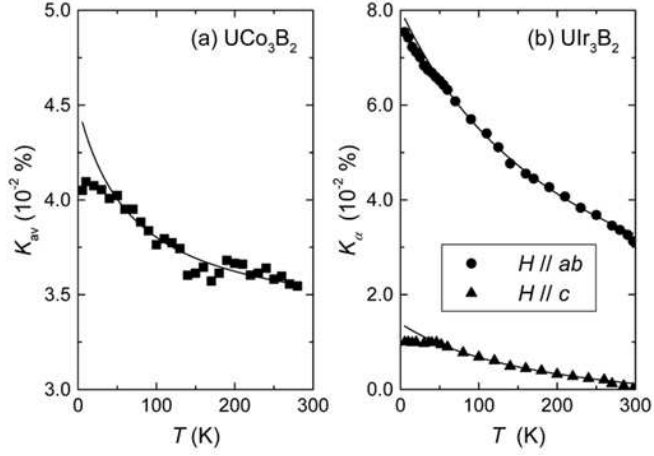}
\caption{\label{K}
The Knight shift vs. temperature plot of (a) UCo$_3$B$_2$ and (b) UIr$_3$B$_2$.
The solid lines are the results of Curie-Weiss fit above 50 K.
}
\end{figure}

Fig. \ref{K} shows the $T$-variation of Knight shift of UCo$_3$B$_2$ ($K_{\textrm{av}}$) and UIr$_3$B$_2$ ($K_{ab}$, $K_{c}$).
Both Knight shift gradually arise with decreasing $T$, whereas these show the saturation tendency below $T^{*}\simeq 50$ K;
$K_{c}$ and $K_\textrm{av}$ display the $T$-independent behavior, but $K_{ab}$ deviate slightly from the continuous variance.
Similar crossover behaviors are also seen in nuclear spin-lattice relaxation rate as described later, and
the obtained results mean that
two different class of magnetic state discriminated at $T^*$ is realized in both U$M_{3}$B$_2$ compounds.

\begin{figure}
\includegraphics{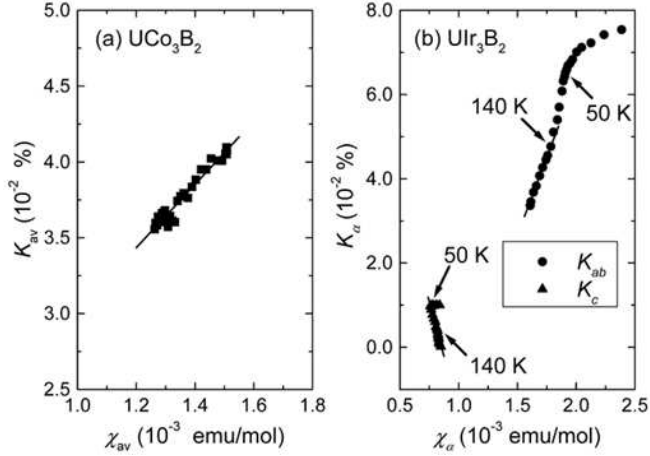}
\caption{\label{K-x}
The Knight shift vs. susceptibility plot of (a) UCo$_3$B$_2$ and (b) UIr$_3$B$_2$.
The solid lines are the linear fit to evaluate the hyperfine coupling constant.
}
\end{figure}

To evaluate the amount of hyperfine interactions between $^{11}$B nuclei and the conduction electrons,
we estimate hyperfine coupling constant ($A_{\textrm{av}}$ for UCo$_3$B$_2$ and $A_{\alpha}$ for UIr$_3$B$_2$)
from the Knight shift vs. susceptibility ($K$-$\chi$) plots.
The results are shown in Fig. \ref{K-x}.
The $K$-$\chi$ of UCo$_3$B$_2$ is straight in all $T$-range,
and $A_\textrm{av}$ derived from the relation $K_{\textrm{av}}=(A_\textrm{av}/N_\textrm{A}\mu_\textrm{B})\chi$ has been 1.2 kOe/$\mu_\textrm{B}$.
On the other hand,
the $K$-$\chi$ in UIr$_3$B$_2$ two times bend at 140 K and 50 K,
which will be clearly seen in $K_{ab}$.
The first anomaly at 140 K will originate from the coherent electron correlations.
Actually, calm drop around similar $T$ has been observed in the resistivity measurement.\cite{previous2,previous3,ikeda}
The second anomaly at 50 K will be caused by the ferromagnetic impurities.
According to the magnetization measurement,\cite{ikeda}
the ferromagnetic hysterisis has been observed and it exhibits rapid enhancement in the $T$-variation of susceptibility below 50 K.
The plausible hyperfine coupling constants obtained from the linear part above 140 K
are $A_{ab}=4.1$ kOe/$\mu_\textrm{B}$ and $A_{c}=-5.8$ kOe/$\mu_\textrm{B}$.
The hyperfine coupling in both materials is rather small than other ligand NMR in $f$-electron systems, and
this will imply that valence band in B atoms are far from the Fermi level.
From the former band calculation by an APW method in UIr$_3$B$_2$,\cite{APW}
the density of state at the Fermi level were assumed to mainly come from the U($5f$) and Ir($5d$) electrons, and
these elements plays the dominant role for physical properties in UIr$_3$B$_2$.
As a result,
small hybridization between B($2s$) bands and conduction electrons will be the origin of the weak hyperfine interactions.

\begin{figure}
\includegraphics{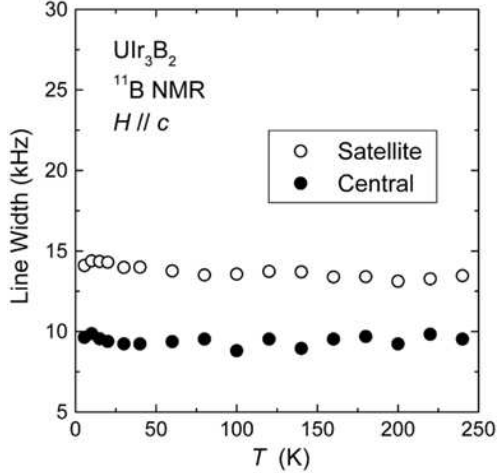}
\caption{\label{width}
The line width vs. temperature plots in UIr$_3$B$_2$.
The magnetic field directs to the $c$-axis, and
no anomaly across the $T^{*}\simeq50$ K is confirmed in this figure.
}
\end{figure}

In UIr$_3$B$_2$,
the $T^{*}$ is similar to the emergent point of ferromagnetic moments.
To verify that the Knight shift (and the nuclear spin-lattice relaxation) anomaly is extrinsic or not,
we checked the $T$-variation of spectrum line width of $^{11}$B NMR.
If the Knight shift anomaly originates from the hyperfine field from the magnetic impurity,
the line width should be also broadened.
The $T$-variation of line width obtained from the central and satellite resonance
along the $c$-axis is displayed in Fig. \ref{width}.
Between 5.0 K and 250 K,
the observed line width is $T$-invariance, and no anomaly across $T^{*}$ could not be confirmed.
Thus,
the Knight shift (and the nuclear spin-lattice relaxation rate) anomaly around the $T^{*}$ is assumed to be intrinsic.

For the qualitative estimation of the magnetic state above the $T^{*}$,
the Curie-Weiss fit of the Knight shift has been performed in both U$M_3$B$_2$ compounds.
The applied fitting function is $K_{\textrm{av},\alpha}=C_{\textrm{av},\alpha}/(T+\theta_{\textrm{av},\alpha})+K_{\textrm{vv}}$,
where $K_{\textrm{vv}}$ is the $T$-independent van Vleck part in the Knight shift.
The results are shown in Fig. \ref{K} as the solid lines.
The obtained Weiss-$T$s are $\theta_{\textrm{av}}=-70$ K, $\theta_{ab}=-241$ K and $\theta_{c}=-189$ K,
and all of these are negative.
Although a simple consideration from the localized picture associates with the antiferromagnetic correlations from the negative Weiss-$T$,
another magnetic state will be claimed by the comparison between the Knight shift and the nuclear spin-lattice relaxation rate.
The detailed analysis will appear in next subsection.
From the estimated Curie constant $C_{\textrm{av},\alpha}$ and the formerly obtained $A_{\textrm{av},\alpha}$,
we calculate effective magnetic moments of U atoms.
The results have been $\mu_\textrm{av}^{\textrm{eff}}=0.55\mu_{\textrm{B}}$, $\mu_{ab}^{\textrm{eff}}=1.49\mu_{\textrm{B}}$ and
$\mu_{c}^{\textrm{eff}}=0.55\mu_{\textrm{B}}$.
Since the full moment of U$^{3+}$ and U$^{4+}$ are in the order of 3.2 $\mu_{\textrm{B}}$,
the estimated magnitudes of magnetic moment imply the itinerancy of $5f$-electron even in the high-$T$ state.

\subsection{\label{NSLR}Nuclear spin-lattice relaxation rates}

From the results of $K_{\textrm{av},\alpha}$ and $[T_1T]_{\textrm{av},\alpha}^{-1}$,
the $T$-variations of $\chi_{q,\omega}$ are discussed in this subsection.
In general,
$K_{\textrm{av},\alpha}$ is in proportion with the uniform susceptibility equivalent to the longitudinal Re[$\chi_{q,\omega}$] at $q,\omega=0$.
On the other hand,
$[T_1T]_{\textrm{av},\alpha}^{-1}$ probes the transverse Im[$\chi_{q,\omega}$] in wide $q$-range, expressed as,\cite{T1expression}
\begin{equation}
[T_1T]^{-1}=\frac{2 \gamma_n^2 k_B}{(\gamma_e \hbar)^2}\sum_qA_qA_{-q}\frac{\textrm{Im}[\chi_{q,\omega_0}^{\perp}]}{\omega_0},
\label{T1T}
\end{equation}
where $A_q$ is the hyperfine coupling constant in $q$-space, and $\omega_0$ is the Larmor frequency of applied rf pulses.
In UIr$_3$B$_2$,
the NMR measurement has been performed with a single crystal,
and consequently
Eqs.~\ref{T1T} means that $[T_1T]_{c}^{-1}$ relates to the $ab$-plane components of Im[$\chi_{q,\omega}$], whereas
$[T_1T]_{ab}^{-1}$ does to the $ab$- and also $c$-axis components.
To gain the knowledge about the anisotropic properties of $\chi_{q,\omega}$,
the next reductions of $[T_1T]_{\alpha}^{-1}$ described as,
\begin{align}
R_{ab} &= [T_1T]^{-1}_{c}, \label{R_ab}\\
R_{c}  &= 2[T_1T]^{-1}_{ab}-[T_1T]^{-1}_{c}, \label{R_c}
\end{align}
have been performed.
As a result,
$R_\alpha$ signifies the longitudinal Im[$\chi_{q,\omega}$] along the $\alpha$-direction,
and the anisotropic properties of magnetic correlations can be discussed from the experimental point of view.
In UCo$_3$B$_2$, meanwhile,
the NMR measurement has been carried out with the powder sample, therefore,
$[T_1T]_{\textrm{av}}^{-1}$ means the spatial average of the nuclear spin-lattice relaxation rates, and
$[T_1T]_{\textrm{av}}^{-1}$ is simply replaced as $R_{\textrm{av}}$.

\begin{figure}
\includegraphics{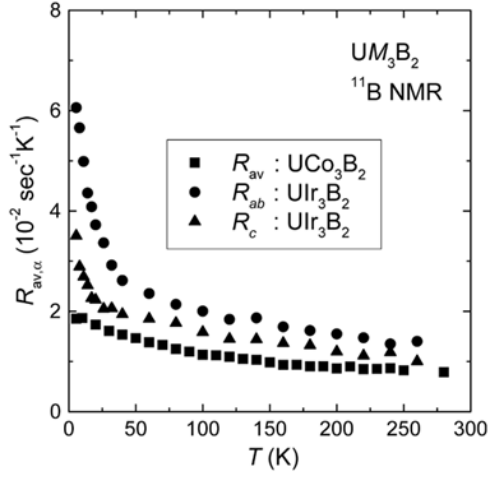}
\caption{\label{R}
The temperature variations of $R_{\textrm{av},\alpha}$.
The means of the squares, circles and triangles are displayed in the explanatory notes.
}
\end{figure}

Fig. \ref{R} shows the $T$-variation of $R_{\textrm{av},\alpha}$.
With decreasing $T$,
the steadily enhancement of $R_{\textrm{av},\alpha}$ not to be the $T$-indepenent Korringa relations in metallic compounds was found.
This result suggests that the magnetic excitations underneath the itinerant Pauli paramagnetism exist in both U$M_{3}$B$_{2}$ compounds.
As mentioned in the former,
the magnetic state crossover in low-$T$ state is also confirmed from the rapid enhancement of $R_{\textrm{av},\alpha}$.
This behavior stands out especially in UIr$_3$B$_2$, and
the rather strong magnetic correlations in UIr$_3$B$_2$ than UCo$_3$B$_2$ will be suspected.

\begin{figure}
\includegraphics{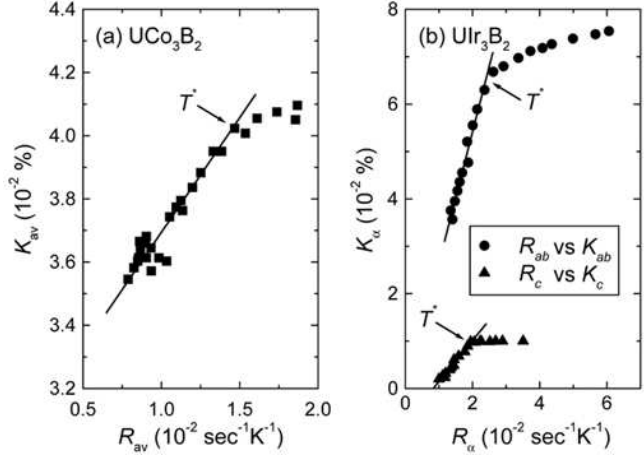}
\caption{\label{K-R}
The $K_{\textrm{av},\alpha}$ vs. $R_{\textrm{av},\alpha}$ plots of (a) UCo$_3$B$_2$ and (b) UIr$_3$B$_2$.
With increasing $K_{\textrm{av},\alpha}$ and $R_{\textrm{av},\alpha}$,
$T$ goes down in this plots.
The solid lines are the guide to eye for the linear relations above $T^{*}\simeq50$ K.
}
\end{figure}

To derive the characteristics of magnetic correlations,
we compared the $T$-variations of $K_{\textrm{av},\alpha}$ with $R_{\textrm{av},\alpha}$.
Fig. \ref{K-R} shows the $K_{\textrm{av},\alpha}$ vs. $R_{\textrm{av},\alpha}$ plots as $T$-implicit.
Above $T^{*}$,
the good linear relations between $K_{\textrm{av},\alpha}$ and $R_{\textrm{av},\alpha}$ has been found in both U$M_3$B$_2$;
in other words,
this result expresses that the $K_{\textrm{av},\alpha}$ and the $R_{\textrm{av},\alpha}$ are proportional to one another.
As described in Eqs. \ref{T1T},
$R_{\textrm{av},\alpha}$ probes the uniform susceptibility ($\chi_{0,0}$) and also the finite-$q$ components of $\chi_{q,\omega}$,
so that the staggered parts in $\chi_{q,\omega}$ should be zero
in order that $R_{\textrm{av},\alpha}$ follows on the same $T$-variations of $K_{\textrm{av},\alpha}$.
The development of $q=0$ spin fluctuations means the ferromagnetic correlations, and
it is interesting that UCo$_3$B$_2$ and UIr$_3$B$_2$ have the same feature in common.
In the itinerant systems with ferromagnetic exchange,
the similar $T$-variance of $K_{\textrm{av},\alpha}$ and $R_{\textrm{av},\alpha}$ sometimes occurs.
The typical substances will be the Laves phase compunds,\cite{Laves} and
the theoretical explanation has been already performed in the framework of self-consistent renormalization theory.\cite{FM-SCR}

Below $T^{*}$,
the $K_{\textrm{av},\alpha}$ vs. $R_{\textrm{av},\alpha}$ plots abruptly bend in both U$M_3$B$_2$ as shown in Fig. \ref{K-R}.
The breakdown of the scaling law results from
the incompatible behaviors between $K_{\textrm{av},\alpha}$ and $R_{\textrm{av},\alpha}$; namely,
$K_{\textrm{av},\alpha}$ shows a $T$-invariant tendency, but
$R_{\textrm{av},\alpha}$ exhibits the monotonic increase.
From the same consideration above $T^*$,
the behavior of $K_{\textrm{av},\alpha}$ suggests that the ferromagnetic component prescribed by $\chi_{0,0}$ almost saturates.
On the other hand,
the unexpected enhancement of $R_{\textrm{av},\alpha}$ against the $T$-invariant tendency of $K_{\textrm{av},\alpha}$ is interpreted as
the appearance of the antiferromagnetic correlations attributed from the finite-$q$ dispersions emerged in $\chi_{q,\omega}$.
The remarkable feature below $T^{*}$ is that the spectral weight of $\chi_{0,0}$ still remains,
because the spin part of $K_{\textrm{av},\alpha}$ does not go down to zero with decreasing $T$.
Furthermore,
$\chi_{0,0}$ seems to form the complicated dispersions with the finite-$q$ components of $\chi_{q,\omega}$.
This assumption suggests that
the ferromagnetic and antiferromagnetic correlations is present at the same time in low-$T$ regime.

It is noteworthy that UCo$_3$B$_2$ and UIr$_3$B$_2$ show the similar magnetic properties;
the ferromagnetic correlations develop above $T^{*}$, and
ferromagnetic and antiferromagnetic correlations simultaneously emerge below $T^{*}$.
From the comparison of each compounds,
it is mentioned that
the structural lattice is almost isomorphic.
In addition,
the valence state is assumed to also closely resemble, because Co and Ir locate at the same column in the periodic table.
Meanwhile,
the $d$-bands formed by the transition metals are different due to the Co-$3d$ and Ir-$5d$ configurations.
The antiferromagnetic correlations infered from $R_{\textrm{av},\alpha}$ have been more emphasized in UIr$_3$B$_2$ than UCo$_3$B$_2$.
This will be caused from the different degree of the hybridization between the $3d$-$5f$ and $5d$-$5f$ state.
On the other hand,
similar magnetic properties will originate from the structural similarity in each compounds.
As described in the Sec. \ref{introduction},
$Ln$Rh$_3$B$_2$ has a quasi-one dimensional electrons band\cite{CeRh3B2,PrRh3B2}
because of the highly compressed structure along the $c$-axis. 
One plausible exposition by a structural clue will be a nesting effect of the ferromagnetic band.
In such a case, ferromagnetic correlations above $T^{*}$ will be caused by the intra-band spin exchange.
Meanwhile, flat regions of Fermi surface may lead to substantial nesting effects with a large contact area, which may in turn
give rise to antiferromagnetic correlations with a finite-$q$ below $T^{*}$.
Although the experimental band investigations in U$M_3$B$_2$ have not been done yet,
an implication for the one-dimensionality is the good conductivity along the $c$-axis observed in UIr$_3$B$_2$,\cite{ikeda} and
this matter of fact can be associated with a flat part in Fermi surface along the $ab$-plane.
For the further investigation of the Fermi surface geometry,
the de Hass-van Alphen and/or the angle resolved photo emission experiment are needed
to understand the magnetic correlations obtained by NMR.

\section{\label{conclusion}Conclusion}
We have performed $^{11}$B NMR measurements on the powdered UCo$_3$B$_2$ and the single crystal UIr$_3$B$_2$.
From the NMR spectrum in UIr$_3$B$_2$ observed in various crystal axes,
crystallographically single B site was identified, and
the ligands arrangements on the two different B sites face the same way by the $60^{\circ}$ or the $120^{\circ}$ rotation around the $c$-axis.
This has been the consistent result with the recently proposed structure of UIr$_3$B$_2$
examined by the X-ray diffraction measurement\cite{yamaura} using the same piece of ingot.
From the $K$-$\chi$ plots,
the magnitude of the hyperfine coupling constant was estimated, and then
$A_{\textrm{av}}=1.2$ kOe/$\mu_{\textrm{B}}$ for UCo$_3$B$_2$, $A_{ab}=4.1$ kOe/$\mu_{\textrm{B}}$ and $A_{c}=5.8$ kOe/$\mu_{\textrm{B}}$
for UIr$_3$B$_2$ have been obtained.
The $T$-variations of $K_{\textrm{av},\alpha}$ and $R_{\textrm{av},\alpha}$ in both U$M_3$B$_2$ have been revealed
the magnetic state crossover at the same point $T^{*}\simeq50$ K.
The development of $\chi_{q,\omega}$ evaluated from $K_{\textrm{av},\alpha}$ and $R_{\textrm{av},\alpha}$ have
clarified the similar characteristics of the magnetic correlations in both U$M_3$B$_2$.
Above $T^*$,
the ferromagnetic correlations evidenced from the development of $\chi_{0,0}$ was found.
On the other hand, below $T^{*}$,
the antiferromagnetic correlations suggested from the finite-$q$ components in $\chi_{q,\omega}$ abruptly emerges.
Since the ferromagnetic correlations does not show the tendency of the reduction below $T^{*}$,
the ferromagnetic and the antiferromagnetic correlations seem to exist simultaneously in low-$T$ phase.
To understand the magnetic properties observed in U$M_3$B$_2$,
the discussion based on the structural origin have been performed.

\begin{acknowledgments}
The authors would like to thank to Prof. H. Yasuoka and Dr. W. Higemoto for their valuable comments and discussions.
\end{acknowledgments}

\newpage 
\bibliography{apssamp}

\begin{thebibliography}{25}
\expandafter\ifx\csname natexlab\endcsname\relax\def\natexlab#1{#1}\fi
\expandafter\ifx\csname bibnamefont\endcsname\relax
  \def\bibnamefont#1{#1}\fi
\expandafter\ifx\csname bibfnamefont\endcsname\relax
  \def\bibfnamefont#1{#1}\fi
\expandafter\ifx\csname citenamefont\endcsname\relax
  \def\citenamefont#1{#1}\fi
\expandafter\ifx\csname url\endcsname\relax
  \def\url#1{\texttt{#1}}\fi
\expandafter\ifx\csname urlprefix\endcsname\relax\def\urlprefix{URL }\fi
\providecommand{\bibinfo}[2]{#2}
\providecommand{\eprint}[2][]{\url{#2}}

\bibitem[{\citenamefont{Peierls}(1955)}]{peierls}
\bibinfo{author}{\bibfnamefont{R.~E.} \bibnamefont{Peierls}},
  \emph{\bibinfo{title}{Quantum Theory of Solid}} (\bibinfo{publisher}{Oxford
  Univ. Press}, \bibinfo{address}{Oxford, England}, \bibinfo{year}{1955}).

\bibitem[{\citenamefont{Tomonaga}(1950)}]{tomonaga}
\bibinfo{author}{\bibfnamefont{S.}~\bibnamefont{Tomonaga}},
  \bibinfo{journal}{Prog. Theor. Phys.} \textbf{\bibinfo{volume}{5}},
  \bibinfo{pages}{544} (\bibinfo{year}{1950}).

\bibitem[{\citenamefont{Luttinger}(1963)}]{luttingere}
\bibinfo{author}{\bibfnamefont{J.~M.} \bibnamefont{Luttinger}}
  (\bibinfo{year}{1963}).

\bibitem[{\citenamefont{Anderson}(1973)}]{RVB}
\bibinfo{author}{\bibfnamefont{P.~W.} \bibnamefont{Anderson}},
  \bibinfo{journal}{Mater. Res. Bull.} \textbf{\bibinfo{volume}{8}},
  \bibinfo{pages}{153} (\bibinfo{year}{1973}).

\bibitem[{\citenamefont{Shimizu et~al.}(2003)\citenamefont{Shimizu, Miyagawa,
  Kanoda, Maesato, and Saito}}]{RVB-exp1}
\bibinfo{author}{\bibfnamefont{Y.}~\bibnamefont{Shimizu}},
  \bibinfo{author}{\bibfnamefont{K.}~\bibnamefont{Miyagawa}},
  \bibinfo{author}{\bibfnamefont{K.}~\bibnamefont{Kanoda}},
  \bibinfo{author}{\bibfnamefont{M.}~\bibnamefont{Maesato}}, \bibnamefont{and}
  \bibinfo{author}{\bibfnamefont{G.}~\bibnamefont{Saito}},
  \bibinfo{journal}{Phys. Rev. Lett.} \textbf{\bibinfo{volume}{91}},
  \bibinfo{pages}{107001} (\bibinfo{year}{2003}).

\bibitem[{\citenamefont{Okubo et~al.}(2003)\citenamefont{Okubo, Yamada,
  Thamizhavel, Kirita, Inada, Settai, Harima, Takegahara, Galatanu, Yamamoto
  et~al.}}]{CeRh3B2}
\bibinfo{author}{\bibfnamefont{T.}~\bibnamefont{Okubo}},
  \bibinfo{author}{\bibfnamefont{M.}~\bibnamefont{Yamada}},
  \bibinfo{author}{\bibfnamefont{A.}~\bibnamefont{Thamizhavel}},
  \bibinfo{author}{\bibfnamefont{S.}~\bibnamefont{Kirita}},
  \bibinfo{author}{\bibfnamefont{Y.}~\bibnamefont{Inada}},
  \bibinfo{author}{\bibfnamefont{R.}~\bibnamefont{Settai}},
  \bibinfo{author}{\bibfnamefont{H.}~\bibnamefont{Harima}},
  \bibinfo{author}{\bibfnamefont{K.}~\bibnamefont{Takegahara}},
  \bibinfo{author}{\bibfnamefont{A.}~\bibnamefont{Galatanu}},
  \bibinfo{author}{\bibfnamefont{E.}~\bibnamefont{Yamamoto}},
  \bibnamefont{et~al.}, \bibinfo{journal}{J. Phys.: Condens. Matter}
  \textbf{\bibinfo{volume}{15}}, \bibinfo{pages}{L721} (\bibinfo{year}{2003}).

\bibitem[{\citenamefont{Yamada et~al.}(2004)\citenamefont{Yamada, Obiraki,
  Okubo, Shiromoto, Kida, Shiimoto, Kohara, Yamamoto, Honda, Galatanu
  et~al.}}]{PrRh3B2}
\bibinfo{author}{\bibfnamefont{M.}~\bibnamefont{Yamada}},
  \bibinfo{author}{\bibfnamefont{Y.}~\bibnamefont{Obiraki}},
  \bibinfo{author}{\bibfnamefont{T.}~\bibnamefont{Okubo}},
  \bibinfo{author}{\bibfnamefont{T.}~\bibnamefont{Shiromoto}},
  \bibinfo{author}{\bibfnamefont{Y.}~\bibnamefont{Kida}},
  \bibinfo{author}{\bibfnamefont{M.}~\bibnamefont{Shiimoto}},
  \bibinfo{author}{\bibfnamefont{H.}~\bibnamefont{Kohara}},
  \bibinfo{author}{\bibfnamefont{T.}~\bibnamefont{Yamamoto}},
  \bibinfo{author}{\bibfnamefont{D.}~\bibnamefont{Honda}},
  \bibinfo{author}{\bibfnamefont{A.}~\bibnamefont{Galatanu}},
  \bibnamefont{et~al.}, \bibinfo{journal}{J. Phys. Soc. Jpn.}
  \textbf{\bibinfo{volume}{73}}, \bibinfo{pages}{2266} (\bibinfo{year}{2004}).

\bibitem[{\citenamefont{Ku et~al.}(1980)\citenamefont{Ku, Meisner, Acker, and
  Johnston}}]{Struc1}
\bibinfo{author}{\bibfnamefont{H.~C.} \bibnamefont{Ku}},
  \bibinfo{author}{\bibfnamefont{G.~P.} \bibnamefont{Meisner}},
  \bibinfo{author}{\bibfnamefont{F.}~\bibnamefont{Acker}}, \bibnamefont{and}
  \bibinfo{author}{\bibfnamefont{D.~C.} \bibnamefont{Johnston}},
  \bibinfo{journal}{Solid State Commun.} \textbf{\bibinfo{volume}{35}},
  \bibinfo{pages}{91} (\bibinfo{year}{1980}).

\bibitem[{\citenamefont{Langen et~al.}(1987)\citenamefont{Langen, Jackel,
  Schlabitz, Veit, and Wlhlleben}}]{Struc2}
\bibinfo{author}{\bibfnamefont{J.}~\bibnamefont{Langen}},
  \bibinfo{author}{\bibfnamefont{G.}~\bibnamefont{Jackel}},
  \bibinfo{author}{\bibfnamefont{G.}~\bibnamefont{Schlabitz}},
  \bibinfo{author}{\bibfnamefont{W.}~\bibnamefont{Veit}}, \bibnamefont{and}
  \bibinfo{author}{\bibfnamefont{D.}~\bibnamefont{Wlhlleben}},
  \bibinfo{journal}{Solid State Commun.} \textbf{\bibinfo{volume}{64}},
  \bibinfo{pages}{169} (\bibinfo{year}{1987}).

\bibitem[{\citenamefont{Misemer et~al.}(1984)\citenamefont{Misemer, Auluck,
  Kobayashi, and Harmon}}]{Struc3}
\bibinfo{author}{\bibfnamefont{D.~K.} \bibnamefont{Misemer}},
  \bibinfo{author}{\bibfnamefont{S.}~\bibnamefont{Auluck}},
  \bibinfo{author}{\bibfnamefont{S.~I.} \bibnamefont{Kobayashi}},
  \bibnamefont{and} \bibinfo{author}{\bibfnamefont{B.~N.}
  \bibnamefont{Harmon}}, \bibinfo{journal}{Solid State Commun.}
  \textbf{\bibinfo{volume}{52}}, \bibinfo{pages}{955} (\bibinfo{year}{1984}).

\bibitem[{\citenamefont{Yamaura et~al.}()}]{yamaura}
\bibinfo{author}{\bibfnamefont{J.}~\bibnamefont{Yamaura}} \bibnamefont{et~al.},
  \bibinfo{note}{recent X-ray results could not assign the scattering profile
  with the formerly reported P6/mmm symmetry, and complex but similar structure
  to CeCo$_3$B$_2$-type has been suggested. The detailed description will
  appear in separate paper.}

\bibitem[{\citenamefont{Yang et~al.}(1985{\natexlab{a}})\citenamefont{Yang,
  Torikachvili, Maple, Ku, Pate, Lindau, and Allen}}]{previous1}
\bibinfo{author}{\bibfnamefont{K.~N.} \bibnamefont{Yang}},
  \bibinfo{author}{\bibfnamefont{M.~S.} \bibnamefont{Torikachvili}},
  \bibinfo{author}{\bibfnamefont{M.~B.} \bibnamefont{Maple}},
  \bibinfo{author}{\bibfnamefont{H.~C.} \bibnamefont{Ku}},
  \bibinfo{author}{\bibfnamefont{B.~B.} \bibnamefont{Pate}},
  \bibinfo{author}{\bibfnamefont{I.}~\bibnamefont{Lindau}}, \bibnamefont{and}
  \bibinfo{author}{\bibfnamefont{J.~W.} \bibnamefont{Allen}},
  \bibinfo{journal}{J. Magn. Magn. Mater.} \textbf{\bibinfo{volume}{47\&48}},
  \bibinfo{pages}{558} (\bibinfo{year}{1985}{\natexlab{a}}).

\bibitem[{\citenamefont{Yang et~al.}(1984)\citenamefont{Yang, Torikachvili,
  Maple, and Ku}}]{previous2}
\bibinfo{author}{\bibfnamefont{K.~N.} \bibnamefont{Yang}},
  \bibinfo{author}{\bibfnamefont{M.~S.} \bibnamefont{Torikachvili}},
  \bibinfo{author}{\bibfnamefont{M.~B.} \bibnamefont{Maple}}, \bibnamefont{and}
  \bibinfo{author}{\bibfnamefont{H.~C.} \bibnamefont{Ku}}, \bibinfo{journal}{J.
  Low Temp. Phys.} \textbf{\bibinfo{volume}{56}}, \bibinfo{pages}{601}
  (\bibinfo{year}{1984}).

\bibitem[{\citenamefont{Yang et~al.}(1985{\natexlab{b}})\citenamefont{Yang,
  Torikachvili, Maple, and Ku}}]{previous3}
\bibinfo{author}{\bibfnamefont{K.~N.} \bibnamefont{Yang}},
  \bibinfo{author}{\bibfnamefont{M.~S.} \bibnamefont{Torikachvili}},
  \bibinfo{author}{\bibfnamefont{M.~B.} \bibnamefont{Maple}}, \bibnamefont{and}
  \bibinfo{author}{\bibfnamefont{H.~C.} \bibnamefont{Ku}}, \bibinfo{journal}{J.
  Appl. Phys.} \textbf{\bibinfo{volume}{57}}, \bibinfo{pages}{3140}
  (\bibinfo{year}{1985}{\natexlab{b}}).

\bibitem[{\citenamefont{Ikeda et~al.}()}]{ikeda}
\bibinfo{author}{\bibfnamefont{S.}~\bibnamefont{Ikeda}} \bibnamefont{et~al.},
  \bibinfo{note}{(\textit{unpublished})}.

\bibitem[{\citenamefont{Fujimoto et~al.}({\natexlab{a}})\citenamefont{Fujimoto,
  Sakai, Tokunaga, Kambe, Walstedt, Ikeda, Matsuda, Haga, and
  \={O}nuki}}]{fujimoto1}
\bibinfo{author}{\bibfnamefont{T.}~\bibnamefont{Fujimoto}},
  \bibinfo{author}{\bibfnamefont{H.}~\bibnamefont{Sakai}},
  \bibinfo{author}{\bibfnamefont{Y.}~\bibnamefont{Tokunaga}},
  \bibinfo{author}{\bibfnamefont{S.}~\bibnamefont{Kambe}},
  \bibinfo{author}{\bibfnamefont{R.~E.} \bibnamefont{Walstedt}},
  \bibinfo{author}{\bibfnamefont{S.}~\bibnamefont{Ikeda}},
  \bibinfo{author}{\bibfnamefont{T.~D.} \bibnamefont{Matsuda}},
  \bibinfo{author}{\bibfnamefont{Y.}~\bibnamefont{Haga}}, \bibnamefont{and}
  \bibinfo{author}{\bibfnamefont{Y.}~\bibnamefont{\={O}nuki}},
  \bibinfo{note}{physica B at press}.

\bibitem[{\citenamefont{Fujimoto et~al.}({\natexlab{b}})\citenamefont{Fujimoto,
  Sakai, Tokunaga, Kambe, Walstedt, Ikeda, Matsuda, Haga, and
  \={O}nuki}}]{fujimoto2}
\bibinfo{author}{\bibfnamefont{T.}~\bibnamefont{Fujimoto}},
  \bibinfo{author}{\bibfnamefont{H.}~\bibnamefont{Sakai}},
  \bibinfo{author}{\bibfnamefont{Y.}~\bibnamefont{Tokunaga}},
  \bibinfo{author}{\bibfnamefont{S.}~\bibnamefont{Kambe}},
  \bibinfo{author}{\bibfnamefont{R.~E.} \bibnamefont{Walstedt}},
  \bibinfo{author}{\bibfnamefont{S.}~\bibnamefont{Ikeda}},
  \bibinfo{author}{\bibfnamefont{T.~D.} \bibnamefont{Matsuda}},
  \bibinfo{author}{\bibfnamefont{Y.}~\bibnamefont{Haga}}, \bibnamefont{and}
  \bibinfo{author}{\bibfnamefont{Y.}~\bibnamefont{\={O}nuki}},
  \bibinfo{note}{j. Phys. Soc. Jpn. Suppl. at press}.

\bibitem[{twi()}]{twin}
\bibinfo{note}{X-ray analisis by J. Yamaura \textit{et al}. has been suggested
  the twinning in a growned UIr$_3$B$_2$. However, obtained NMR spectrum could
  be assigned by the single phase. This cause is that NMR spectrum observed
  from each domains completely overlap with each other.}

\bibitem[{\citenamefont{Narath}(1967)}]{Narath}
\bibinfo{author}{\bibfnamefont{A.}~\bibnamefont{Narath}},
  \bibinfo{journal}{Phys. Rev.} \textbf{\bibinfo{volume}{162}},
  \bibinfo{pages}{320} (\bibinfo{year}{1967}).

\bibitem[{\citenamefont{Stauss}(1964)}]{W}
\bibinfo{author}{\bibfnamefont{G.~H.} \bibnamefont{Stauss}},
  \bibinfo{journal}{J. Chem. Phys.} \textbf{\bibinfo{volume}{40}},
  \bibinfo{pages}{1988} (\bibinfo{year}{1964}).

\bibitem[{\citenamefont{Bloembergen and Rowland}(1953)}]{PP}
\bibinfo{author}{\bibfnamefont{N.}~\bibnamefont{Bloembergen}} \bibnamefont{and}
  \bibinfo{author}{\bibfnamefont{T.~J.} \bibnamefont{Rowland}},
  \bibinfo{journal}{Acta Metall.} \textbf{\bibinfo{volume}{1}},
  \bibinfo{pages}{731} (\bibinfo{year}{1953}).

\bibitem[{\citenamefont{Kasaya et~al.}(1990)\citenamefont{Kasaya, Iga, Katoh,
  Takegahara, and Kasuya}}]{APW}
\bibinfo{author}{\bibfnamefont{M.}~\bibnamefont{Kasaya}},
  \bibinfo{author}{\bibfnamefont{F.}~\bibnamefont{Iga}},
  \bibinfo{author}{\bibfnamefont{K.}~\bibnamefont{Katoh}},
  \bibinfo{author}{\bibfnamefont{K.}~\bibnamefont{Takegahara}},
  \bibnamefont{and} \bibinfo{author}{\bibfnamefont{T.}~\bibnamefont{Kasuya}},
  \bibinfo{journal}{J. Magn. Magn. Mater.} \textbf{\bibinfo{volume}{90\&91}},
  \bibinfo{pages}{521} (\bibinfo{year}{1990}).

\bibitem[{\citenamefont{Moriya}(1963)}]{T1expression}
\bibinfo{author}{\bibfnamefont{T.}~\bibnamefont{Moriya}}, \bibinfo{journal}{J.
  Phys. Soc. Jpn.} \textbf{\bibinfo{volume}{18}}, \bibinfo{pages}{516}
  (\bibinfo{year}{1963}).

\bibitem[{\citenamefont{Yoshimura et~al.}(1987)\citenamefont{Yoshimura,
  Fukamichi, Yasuoka, and Mekata}}]{Laves}
\bibinfo{author}{\bibfnamefont{K.}~\bibnamefont{Yoshimura}},
  \bibinfo{author}{\bibfnamefont{K.}~\bibnamefont{Fukamichi}},
  \bibinfo{author}{\bibfnamefont{H.}~\bibnamefont{Yasuoka}}, \bibnamefont{and}
  \bibinfo{author}{\bibfnamefont{M.}~\bibnamefont{Mekata}}
  (\bibinfo{year}{1987}).

\bibitem[{\citenamefont{Moriya and Ueda}(1974)}]{FM-SCR}
\bibinfo{author}{\bibfnamefont{T.}~\bibnamefont{Moriya}} \bibnamefont{and}
  \bibinfo{author}{\bibfnamefont{T.}~\bibnamefont{Ueda}},
  \bibinfo{journal}{Solid State Commun.} \textbf{\bibinfo{volume}{15}},
  \bibinfo{pages}{169} (\bibinfo{year}{1974}).

\end{thebibliography}

\end{document}